\documentclass[aps,prl,twocolumn,tightenlines,epsfig]{revtex4}

\usepackage {graphicx}

\usepackage {graphics}

\usepackage {amsmath}

\usepackage {amsfonts}

\usepackage {amssymb}

\usepackage {mathrsfs}

\usepackage{color}

\usepackage{epstopdf}

\newcommand {\be}{\begin{eqnarray}}

\newcommand {\ee}{\end{eqnarray}}

\begin{document}

\title {Instanton sector of correlated electron systems as the origin of populated pseudo-gap and flat ``band'' behavior: analytic solution}

\author {S.I.\ Mukhin}

\email {sergeimoscow@online.ru}

\affiliation {Theoretical Physics Department, Moscow Institute for Steel \& Alloys, Moscow, Russia}

%\sqrt{}

\date {\today}

\begin{abstract}
Finite temperature instantons between meta-stable vacua of correlated electronic system are solved analytically for quasi one-dimensional Hubbard model. The instantons produce dynamic symmetry breaking and connect metallic state with the dual vacua: superconducting (SC) and spin-density wave (SDW) states. The instantons spread along the Matsubara's imaginary time and possess the structure similar to the coordinate-space solitonic lattices previously discovered in quasi one-dimensional Peierls model. On the microscopic level the inter-vacua excursion is described by mutual transformations between the ``resonating quartets'' of the couples of electron-hole and Cooper pairs. Spectral properties of the electrons in the ``instantonic crystal'' reveal pseudo-gap (PG) behavior, with finite fermionic density of states in the center of the PG and ``flat-band'' outside of it. Analytically derived inverse temperature scaling of the pseudo-gap and the densities of the SC and SDW condensates is discussed in the context of ARPES and STM data in high-T$_c$ cuprates. 
\end{abstract}

\maketitle

It is for long time now that competing orders are considered as one of the core properties of underdoped high-T$_c$ cuprates \cite{coex},\cite{matv}. Nevertheless, the nature of the non-superconducting order remains a mystery. Especially challenging is the pseudo-gap (PG) phenomenon observed in high-T$_c$ cuprates \cite{pg}. Here we study a novel possibility, which exploits a well known concept of high-energy physics, an instanton, that connects vacua of the many-body system  possessing different symmetries \cite{polyakov}. The instanton sector of the Hilbert space of the many-body system is known to possess dynamically generated masses of the excitations in the system \cite{polyakov}. We find that this mechanism in the correlated electron system has interesting peculiarity: the ``mass'' is acquired by fermionic Landau quasi-particles inside the pseudo-gap, but neither right in its center, nor far away from the PG region of the spectrum. In the center of the pseudo-gap the quasi-particle dispersion remains linear in momentum (``relativistic''), but acquires just strong enhancement of the Fermi velocity. Outside the PG region the spectrum remains not renormalized. As a result, in the cross-over region a ``flat band'' forms in the fermionic excitation spectrum, see Figs. 1,2, with a weaker dependence on momentum. We demonstrate these properties using a (quasi) one-dimensional model of electron system, which is especially convenient tool for understanding the major effects that occur, since it proves to be analytically solvable and, hence, provides in-depths information. We also discuss predicted measurable characteristics of the electronic system that undergoes symmetry excursions into the (dual) meta-stable vacua described respectively with superconducting (SC) and spin-density wave (SDW) orders. These vacua are chosen as the most relevant ones since ARPES, STM and INS experiments \cite{pg,shen,argon,stm,barisic} reveal PG behavior in the anti-nodal region of the Fermi surface of doped high-T$_c$ cuprates with coexisting magnetic and superconducting fluctuations well above the superconducting transition temperature even in the optimally doped compounds. From our present point of view, supporting evidence for the discussed instanton scenario also comes from the spectral structure of the magnetic fluctuations measured in doped high-T$_c$ cuprates \cite{aeppli} near the anti-ferromagnetic points of the Brillouin zone. The structure reveals ``rod-like'' behavior of the intensity as function of frequency. This, when Fourier-transformed back to time, points to strong unharmonicity of the magnetic fluctuations, thus reminding an ``instanton hair-comb'' rather than commonly considered RPA-like sinusoidal fluctuations around an equilibrium. A complete theory should describe transitions between different symmetries representing the $SO(5)$ group:  charge- spin- and superconducting orders, relevant to correlated electronic system. Nevertheless, we believe that SC and SDW orders could be less coupled to the lattice degrees of freedom (due to charge transfer in cuprates) than CDW order, thus forming most probable end-points for the ``fast'' instanton routes, of relevance for the $10\div 100$ meV scale, where the ``rod-like'' spectral intensity of the magnetic fluctuations is observed \cite{aeppli}. 
As we address behavior at finite temperatures, an instanton propagates in the Matsubara's imaginary time \cite{agd} and is contained in the partition function $Z$ of the quasi one-dimensional many-body system \cite{polyakov,NHD}:
\be
\label{Z}
&&Z=\int{\cal{D}}\bar{\Psi}{\cal{D}}\Psi{\cal{D}}{M}{\cal{D}}\Delta exp{\left\{-(S_{F}+S_{eF})\right\}}\\
&&S_{eF}=\int\int_0^{\beta=1/T}{dxd\tau  \left[\bar{\Psi}\partial_\tau\Psi+H_{eF}\right]}\\
\label{SeF}
&&S_{F}=\int\int_0^{\beta=1/T}{dx d\tau\left[\dfrac{M^2}{U}+\dfrac{\Delta^2}{g}\right]}.
\label{SF}
\ee
\noindent Here we put the Boltzman's and Plank's constants to be unity ($k_B=\hbar=1$). Hamiltonian of the interacting electrons in the decoupling Hubbard-Stratanovich (HS) fields ${M}(\tau,x)$ and $\Delta(\tau,x)$ has the form (compare \cite{matv}):
\be
&&H_{eF}=\bar{\Psi}_{\sigma}\left(-i\dfrac{\partial}{\partial x}\right)\hat{{\sigma}_z}\Psi_{\sigma}+M(\tau,x)\bar{\Psi}_{\sigma}\hat{\sigma}_+\Psi_{\sigma}+\nonumber\\
&&\bar{M}(\tau,x)\bar{\Psi}_{\sigma}\hat{\sigma}_-\Psi_{\sigma}+\Delta(\tau,x)(-\bar{\Psi}_{+,\uparrow}\bar{\Psi}_{-,\downarrow}+\bar{\Psi}_{-,\uparrow}\bar{\Psi}_{+,\downarrow})+\nonumber\\
&&\bar{\Delta}(\tau,x)(-{\Psi}_{-,\downarrow}{\Psi}_{+,\uparrow}+{\Psi}_{+,\downarrow}{\Psi}_{-,\uparrow}).
\label{ZH}
\ee
\noindent
here spin indices are either indicated by up and down arrows or in the summs by index $\sigma$. The Pauli's matrices $\hat{\sigma}$ are defined on the pseudo-spin $1/2$ space with $+1/2$ and $-1/2$ projections distinguishing between right and left moving electrons in the one-dimensional system with coordinate $x$. The operators $\bar{\Psi}$ and $\Psi$ create and annihilate fermions and are taken in the Heisenberg's representation with the Matsubara's imaginary ``time'' $\tau$. Partition function in Eq. (\ref{Z}) contains only (fermi)bosonic fields that obey (anti)periodic conditions, with the temperature $T$ defining the period $\beta\equiv T^{-1}$ \cite{NHD}:
\be 
\label{BFP}
&&\bar{\Psi}(\tau+\beta),\Psi(\tau+\beta)=-\bar{\Psi}(\tau),-\Psi(\tau);\;\;\\ &&M(\tau+\beta),\Delta(\tau+\beta)=M(\tau),\Delta(\tau).  
\label{BF}
\ee
\noindent
Integration over fermionic (Grassmannian) fields $\bar{\Psi}$ and $\Psi$ in (\ref{Z}) could be (formally) done exactly,
provided that one knows the full spectrum $\{\alpha_n\}$ of the quasi-energies of the related Floquet equation:
\be
\partial_\tau{\Phi}_n+\hat{H}_{eF}(M(\tau),\Delta(\tau)){\Phi}_n=(\alpha_n+i\pi \beta^{-1}){\Phi}_n
\label{FQ}
\ee
\noindent that follows from the Floquet's theorem for the fermionic wave-vunctions in the ``time''-periodoc bosonic fields $M(\tau),\Delta(\tau)$ (analog of the Bloch's theorem for space-periodic potentials)\cite{NHD}:
\be
e^{-i\pi\tau/\beta}{\Psi}_n(\tau)=e^{\alpha_n\tau}\Phi_n(\tau);\;\Phi_n(\tau+\beta)=\Phi_n(\tau)
\label{FQF}
\ee
\noindent Since the indices $\alpha_n$ are known, partition function in Eq. (\ref{Z}) can be expressed as:
\be
Z=\int{\cal{D}}{M}{\cal{D}}\Delta exp{\{-S_{F}\}}\prod_n ch\left(\dfrac{\alpha_n}{2}\right)
\label{ZM}
\ee
\noindent In the approximation studied below, the path-integration in Eq. (\ref{ZM}) is substituted by the saddle-point value in the
Hilbert space of the HS fields $M,\Delta$, that is found from zero variational derivatives equations:
\be
&&\delta_{M(\tau,x),\Delta(\tau,x)}{\left\{ S_F-\sum_n ln\left\{ch\left(\dfrac{\alpha_n}{2}\right)\right\}\right\}}=0
\label{self}
\ee
\noindent While obtaining the self-consistency equations in the explicit form below we were guided by the following mathematical and physical facts and considerations. Firstly, the functional derivatives of the eigen-values $\alpha_n$ with respect to the potentials $M(\tau,x)$ and $\Delta(\tau,x)$ in (\ref{self}) are readily calculated using first-order perturbation theory applied to Eq. (\ref{FQ}):
\be
\partial_{M,\Delta(\tau,x)}\alpha_n={\bar{\Phi}}_n(\tau,x)\{\partial_{M,\Delta}\hat{H}_{eF}(M,\Delta)\}{\Phi}_n(\tau,x)
\label{DFQ}
\ee 
\noindent Secondly,the Hamiltonian matrix $\hat{H}_{eF}(M,\Delta)$, as it follows from Eq. (\ref{ZH}), is a $4\times 4$ matrix defined on the quartets of functions $u_{+},u_{-},v_{+},v_{-}(\tau,x)$. The latter are considered as the amplitudes of the right- and left-moving electrons and holes coupled by SDW and SC fields, which are searched for in the form: $M(\tau,x)=M(\tau) exp\{ik_Fx\}+M^*(\tau)exp\{-ik_Fx\}$ and $\Delta(\tau)$, $k_F$ being the one-dimensional Fermi-momentum.
It proves to be, that in the static (i.e. $\tau$-independent) case, for any arbitrary moduli ratio of the SDW and SC orders: $|M|/|\Delta|$, and for arbitrary phases $\phi_{W}$ and $\phi_S$ of these order parameters, the matrix $\hat{H}_{eF}(M,\Delta)$ can be diagonolized, provided that the $u,v$ amplitudes obey the following conditions \cite{matv}:
\be
v_{\pm}=e^{i(\pm\phi_{W}-\phi_{S})}u_{\mp};\;\;v_{\pm}=-e^{i(\pm\phi_{W}-\phi_{S})}u_{\mp}.
\label{uv}
\ee
\noindent Then, it is logical to assume that the instanton passes between SC ($|M|=0$) and SDW ($|\Delta|=0$) vacua along the trajectory in the Hilbert space curbed by the $\{u_{+},u_{-},v_{+},v_{-}\}$-quartets, that obey conditions (\ref{uv}) at any moment inside the imaginary time period $0\leq\tau\leq\beta$. Under these conditions the Hamiltonian $4\times 4$ matrix is $2\times2$-block diagonal, and equations defining the eigen-functions $\Phi_n(\tau)$ in Eq. (\ref{FQ}) for each $2\times 2$ block read:
\be
\left(\hat{\sigma}_z\partial_{\tau}-\Delta_{\pm}(\tau)\hat{\sigma}_{+}+\bar{\Delta}_{\pm}(\tau)\hat{\sigma}_{-}+vk\hat{1}\right)\vec{u}_k=0
\label{bdg}
\ee
\noindent where $\{\vec{u}_k\}^T=(u_+,u_-)$ are Fourier-components of the slowly varying in space ($x$-coordinate) amplitudes $u_{\pm}(\tau,x)$ of the right/left-moving fermions with the wave-functions: $u=u_+(\tau,x)exp(ik_Fx)+u_-(\tau,x)exp(-ik_Fx)$; $v$- is Fermi velocity. The combined HS fields $\Delta_{\pm}$ are defined as follows:
\be
\Delta_{\pm}(\tau)=(|M(\tau)|\pm |\Delta(\tau)|)e^{i\phi_W}  
\label{dpm}   
\ee
\noindent The choice of the sign $+$ or $-$ inside $\Delta_{\pm}$ distinguishes the two different $2\times 2$-block parts of the subdiagonalized complete $4\times 4$ Hamiltonian matrix. Below we assume $\Delta_{\pm}$ to be real. Then, a unitary transformation: $f_{\pm}=(2)^{-1/2}(u_+\pm u_-)$ transforms Eq. (\ref{bdg}) into:
\be
\left(\partial_{\tau}+vk\hat{\sigma}_{x}-\Delta_{\pm}(\tau)\hat{\sigma}_{z}\right)\vec{f}_k=0; 
\label{bdgf}
\ee
\noindent where $\{\vec{f}_k\}^T=(f_+,f_-)$. In the case of the self-consistent instanton solutions with real fields $\bar{\Delta}_\pm=\Delta_\pm$ found below, the (Matsubara's)time-conjugated functions obey the following symmetry relations:
\be
f_{\pm}(k)\leftrightarrow \bar{f}_{\mp}(-k),
\label{bdgfc}
\ee
\noindent that can be checked directly. Functions $f_{\pm}$ satisfy equations following from Eq. (\ref{bdgf}):
\be
\label{ff}
&&({\partial^2}_{\tau}-Q_{\pm}(\tau)-(vk)^2)f_{\pm}=0;\;\\
&&Q_{\mp}(\tau)={\Delta^2}_{(\pm)}\mp \partial_{\tau}{\Delta_{(\pm)}}
\label{dbdg}
\ee  
\noindent and equations for the conjugated functions $\bar{f}_{\pm}$ can be obtained from Eq. (\ref{ff}) by the simultaneous interchange ${f(k)}_{\pm}\rightarrow {\bar{f}(-k)}_{\mp}$ and $Q_{\pm}(\tau)\rightarrow Q_{\mp}(\tau)$. 
It is obvious from Eqs. (\ref{ff}), (\ref{dbdg}) that they differ from the corresponding equations for the Peierls \cite{br,machida} and one-dimensional Hubbard model \cite{muk} only by the opposite sign in front of the square of the bare dispersion $vk$ and by the interchange of the space variable with the imaginary Matsubara's time $\tau$. This latter interchange defines the period of the $\Delta_{\pm}$-functions to be equal to the inverse temperature $\beta=1/T$,
while in the Peierls model a real-space period is proportional to $2\pi/\epsilon$, i.e. to the doping concentration $\epsilon$ away from the half-filling of the bare electronic band.
Hence, a decrease of the temperature corresponds to ``underdoping''.\\
An explicit expression for the functional derivative of the Floquet indices $\alpha_n$ with respect to combined HS fields $\Delta_{\pm}$ is readily derived from Eq. (\ref{bdgf}) by means of the first-order perturbation theory:
\be
&&\partial_{\Delta_{\pm}(\tau)}\alpha_k=\{\bar{f}_{+}f_{+}-\bar{f}_{-}f_{-}\}_{\tau,k}^{\pm}\equiv\nonumber\\
&&\equiv\{\bar{f}_{+}(k)\bar{f}_{-}(-k)-\bar{f}_{-}(k)\bar{f}_{+}(-k)\}_{\tau}^{\pm}
\label{DFQf}
\ee
\noindent where the last equality is obtained using the symmetry relations in Eq. (\ref{bdgfc}), and superscript $^\pm$ distinguishes solutions of Eq. (\ref{ff}) with the two different HS fields $\Delta_{\pm}$ defined in (\ref{dpm}) . Equation (\ref{DFQf}) can be now substituted into equations (\ref{self}), resulting in the following system of self-consistency equations for the two instanton fields $\Delta_{\pm}$ (linearly related with the SC and SDW fields: $\Delta=1/2(\Delta_+-\Delta_-)$ and $M=1/2(\Delta_++\Delta_-)$):
\be
&&\sum_k th\left(\dfrac{{\alpha_k}^{\pm}}{2}\right){\{\bar{f}_{+}(k)\bar{f}_{-}(-k)-\bar{f}_{-}(k)\bar{f}_{+}(-k)\}_{\tau}^{\pm}}=\nonumber\\
&&={\Delta_+}(\tau)\left( {U}^{-1}+{|g|}^{-1}\right)\pm{\Delta_-}(\tau)\left( {U}^{-1}-{|g|}^{-1}\right)
\label{selfin}
\ee

We found that Eqs. (\ref{ff}), (\ref{dbdg}) and (\ref{selfin}) possess self-consistent {\it{instanton}} solution that can be obtained directly from the self-consistent {\it{soliton}} solution found previously \cite{br,machida}. Thus, the former solution can be expressed via the Weierstrass's $\cal{P}$ and Jacobi's snoidal $\Delta_2 sn(\Delta_1;k_1)$ elliptic functions, that also described solitons in the one-dimensional chains.  Only now the Matsubara's imaginary time plays the role of space-coordinate, hence, leading to the following self-consistent instanton solution of the equations (\ref{ff}), (\ref{dbdg}) and (\ref{selfin}) (compare \cite{br,machida}):
\be
\label{felli}
&&f_{\pm}(\tau,k)=\dfrac{w_{\pm}(t)}{{\cal{N}}^{1/2}} exp\left\{{\displaystyle \mp C(k)\int_0^\tau \phi_{\pm}(t_{\pm}) d{\tau '}}\right\}\\
&&w_{\pm}(t)=\sqrt{e-t_{\pm}};\;\;\phi(t)=\dfrac{1}{e-t_{\pm}};\;\;\\
&&t_{\pm}={\cal{P}}(\tau+\tau_{\pm});\;\;\tau_{\pm}=\omega_{2,3}+\tau_0; \\
&&C(k)=\prod_{i=1}^3\sqrt{(e-e_i)};\;\;e=e_1+(vk)^2
\label{elli}
\ee
\noindent Here ${\cal{N}}=e-\bar{\cal{P}}$ is a normalization constant (constant $\bar{\cal{P}}$ is explicitly expressed below), $e_1>e_2>e_3$ are the three values of the Weierstrass function $\cal{P}$ at its three half-periods $\omega_1$, $\omega_2$ and $\omega_3=-(\omega_1+\omega_2)$, that are related with the parameters $\Delta_{1,2}$ and $k_1$ of the corresponding Jacobi function $\Delta_2 sn(\Delta_1\tau;k_1)$. The latter describes the instantons $\Delta_{\pm}(\tau)$:
\be
&&\Delta_{\pm}=\Delta_2^{\pm}sn\left(\Delta_1^{\pm}\tau;k_1^{\pm}\right);\;e_1^{\pm}=\dfrac{\{\Delta_1^{\pm}\}^2+\{\Delta_1^{\pm}\}^2}{6};\;\nonumber\\
&&e_{2,3}^{\pm}=-\dfrac{e_1^{\pm}}{2}\pm \dfrac{\Delta_1^{\pm} \Delta_2^{\pm}}{2};\;k_1^{\pm}=\dfrac{\Delta_2^{\pm}}{\Delta_1^{\pm}}.
\label{snoi}
\ee 
\noindent It is well known that both elliptic functions $\cal{P}$ and $\Delta_{\pm}=\Delta_2 sn(\Delta_1\tau;k_1)$ share the same period $2\omega_1$ as functions of the real variable $\tau$, provided they are related in the following way \cite{witt}:
\be
2{\cal{P}}(\tau+\tau_{\pm})+e_1=\Delta_{(\pm)}^2\pm\partial_\tau{\Delta_{(\pm)}} 
\label{wesnoi}
\ee
\noindent Hence, the Matsubara's time periodicity conditions (\ref{BF}) lead to the following self-consistency equation (two equations appear if we retain the indices $^{\pm}$ distinguishing between the two instanton fields $\Delta_{\pm}$):
\be
\beta\equiv\dfrac{1}{T}=2\omega_1\equiv\dfrac{2K({\tilde{k}})}{(e_1-e_3)^{1/2}};\;{{\tilde{k}}}^2=\dfrac{e_2-e_3}{e_1-e_3},
\label{seleq1}
\ee
\noindent where $T$ is the temperature, and $K(\tilde{k})$ is the complete elliptic integral of the first kind. The remaining two self-consistency equations follow then from Eq. (\ref{selfin}), that due to Eq. (\ref{felli}) become remarkably simple:
\be
&&2\Delta_{\pm}(\tau)\sum_k th\left(\dfrac{{\alpha_k}^{\pm}}{2}\right)\frac{vk}{e^{\pm}-\bar{{\cal{P}}}^{\pm}}=\nonumber\\
&&={\Delta_+}(\tau)\left( \dfrac{1}{U}+\dfrac{1}{|g|}\right)\pm{\Delta_-}(\tau)\left( \dfrac{1}{U}-\dfrac{1}{|g|}\right)
\label{selfinn}
\ee
\noindent where $\bar{{\cal{P}}}$ is half-period averaged Weierstrass function:
\be
&&\bar{{\cal{P}}}=\int_0^{\omega_1}\dfrac{{\cal{P}}(x+\omega_{2})}{\omega_1}dx=e_1-(e_1-e_3)\dfrac{E(\tilde{k})}{K(\tilde{k})}
\label{weier}
\ee
\noindent where $E(\tilde{k})$ is the complete elliptic integral of the second kind. It is possible now, using definition of the Floquet's indices from Eq. (\ref{FQF}) and solutions for the wave-functions from Eqs. (\ref{felli})-(\ref{elli}), to find an explicit analytical expression for the spectrum of fermion quasi-energies in the presence of ``traversing'' instantons:  
\be
\alpha_k=C(k)\int_0^{\beta=2\omega_1}\dfrac{1}{2{\cal{P}}(\tau+\omega_{2})-e}d{\tau}
\label{spec}
\ee
\noindent The well known properties of the Weierstrass function \cite{witt} and of the complete elliptic integral of the third kind $\Pi(n,{k})$ enable us to find the analytic answer in a compact form (omitting the superscript $^\pm$ for the sake of clarity):
\be
&&\alpha_k=2\epsilon_k\left(\dfrac{1-{\tilde{k}}^2+\epsilon_k^2}{1+\epsilon_k^2}\right)^{1/2}
\Pi\left(\dfrac{{\tilde{k}}^2}{1+\epsilon_k^2},{\tilde{k}}\right)\\
&&\epsilon_k=\dfrac{vk}{2TK({\tilde{k}})}
\label{specpi}
\ee
\noindent This expression is remarkable, since it reveals the pseudo-gap with ``localized energy'' states inside of it, and a ``flat-band'' behavior between the gap edge and the ``free-fermion'' spectrum, see Figs. 1,2. These features remind well known data measured e.g. by ARPES in high-T$_c$ cuprates. The corresponding asymptotic expressions read:

\be
\label{aspec1}
&&\alpha_k\approx \dfrac{vk}{T}\dfrac{O(1)}{k'K(\tilde{k})};\;vk<k'TK(\tilde{k})\\
&&\alpha_k\approx 2ln\left(\dfrac{4TK(\tilde{k})}{vkk'}\right);\;k'<\dfrac{vk}{TK(\tilde{k})}<1\\
&&\alpha_k\approx \dfrac{vk}{T};\;TK(\tilde{k})<vk<\varepsilon_F.
\label{aspec3}
\ee
\noindent where $\varepsilon_F$ is of the order of the bare electron energy cut-off, and $k'^2=1-\tilde{k}^2$, to be found self-consistently below.
\\
Using definition of $e$ from (\ref{elli}) and equation (\ref{weier}) we rewrite self-consistency equations (\ref{selfinn}) in the form:
\be
\Delta_{\pm}(\tau)\gamma_\pm={\Delta_+}(\tau)\delta_{+}\pm{\Delta_-}(\tau)\delta_{-}
\label{selfip}
\ee
\noindent where we introduced notation:
\be
\delta_{\pm}=\left( {U}^{-1}\pm{|g|}^{-1}\right);
\label{n}
\ee
\noindent and the essential coefficients $\gamma_\pm$ are:
\be
\gamma_\pm=2\sum_k th\left(\dfrac{{\alpha_k}^{\pm}}{2}\right)\frac{vk}{(vk)^2+4(T)^2K(\tilde{k}^{\pm})E(\tilde{k}^\pm)}
\label{gammas}
\ee
\noindent We still have choice with respect to the ''initial times'' $\tau_0^{\pm}$ in Eq. (\ref{selfip}) since $\Delta_{\pm}(\tau)=\Delta_2^{\pm} sn(\Delta_1^{\pm}(\tau+\tau_0^{\pm});k_1^{\pm})$. Since the self-consistency equations Eq. (\ref{selfip}) fulfill in each moment of ``time'' $\tau$ there are only two irreducible possibilities: $\tau_0^\pm=0$ and $\tau_0^+-\tau_0^-=\omega_1/\Delta_1$. In the first case $\Delta_\pm(\tau)\propto sn(\Delta_1^{\pm}(\tau);k_1^{\pm})$,
and in the second case $\Delta_\pm(\tau)\propto \pm sn(\Delta_1^{\pm}(\tau);k_1^{\pm})$. The second case proved to have no solution with the pseudo-gap property of the fermionic spectrum. Hence, we consider below the first case, i.e. $\tau_0^{\pm}=0$.
It follows directly from Eqs. (\ref{aspec1})-(\ref{aspec3}) that essential (logarithmic) contribution to the sum in 
Eq. (\ref{gammas}) comes from the momenta interval: $k'TK(\tilde{k})< vk < \varepsilon_F$. This leads to the following result:
\be
\gamma_\pm\approx\dfrac{1}{\varepsilon_F}ln \left({\varepsilon_F}/{2T\sqrt{K(\tilde{k}^{\pm})E(\tilde{k}^\pm)}}\right)
\label{gammasfin}
\ee
\noindent Equations (\ref{aspec1})-(\ref{aspec3}) indicate that the pseudo-gap behavior is present provided that $k'<<1$. The self-consistent solution possessing this property reads:
\be
\label{scsol}
&&\Delta(\tau)=\dfrac{U-|g|}{2U}M(\tau);\\
&&M(\tau)=\Delta_M\left(1-\dfrac{U-|g|}{2U}\right) sn(4K(\tilde{k})T\tau;\tilde{k});\\
\label{msol}
&&\Delta_M\equiv\dfrac{\varepsilon_F^2}{2T}exp\left\{-\dfrac{2\varepsilon_F}{|g|}\right\},\; T<T^*\\
\label{m}
&&k'=4exp\{-\Theta\};\;\Theta=\left(\dfrac{\varepsilon_F}{2T}\right)^2exp\left\{-\dfrac{2\varepsilon_F}{|g|}\right\}\\
\label{kprime}
&&\Delta_M=0,\;T>T^*;\;T^*\propto{\varepsilon_F}exp\left\{-\dfrac{\varepsilon_F}{|g|}\right\}
\label{tstar}
\ee
\noindent Substituting $k'$ from Eq. (\ref{kprime}) into Eqs. (\ref{aspec1})-(\ref{aspec3}) we evaluate the pseudo-gap $\Delta_{PG}$: $\Delta_{PG}\propto\Delta_M\sim 1/T$.
\\
In summary, we see from Eqs. (\ref{scsol})-(\ref{tstar}) that when electron coupling constant $|g|$ is close to on-site repulsion $U$ the co-existing superconducting and spin-density wave fluctuating condensates (instantons) form below the temperature $T^*$ that depends on superconducting coupling $|g|$, with both condensates densities being inverse proportional to temperature. The density of the superconducting condensate $\Delta$ vanishes at the point $|g|=U$. Also, a pseudo-gap in the fermionic energy spectrum exists with the magnitude $\Delta_{PG}$ proportional to the density of the magnetic condensate $\Delta_M$. In the center of the pseudo-gap there are fermionic states with the density of states that decreases exponentially with temperature according to ``quadratic Arrenius'' law $\sim k'=exp\{-const\cdot({\varepsilon_F}/{2T})^2\}$, where $const=exp\{-{2\varepsilon_F}/{|g|}\}$. These in-gap states are also observed experimentally \cite{barisic} in high-T$_c$ cuprates.

\begin{figure}
\begin{center} 
\includegraphics[width=0.45\textwidth]{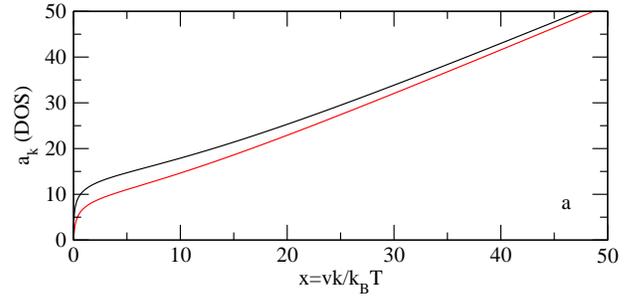}
\end{center}
\caption{Pseudo-energies spectrum for different values of the coupling strength parameter $k'$ as given in Eq. (\ref{kprime}), see text.}
\label{1}
\end{figure}

\noindent
Also in qualitative accord with the ARPES experiments in these compounds \cite{pg,shen,argon} we find at the edge of the pseudo-gap a ``flat band'' in the fermionic dispersion, see Fig. 2.

\begin{figure}
\begin{center} 
\includegraphics[width=0.45\textwidth]{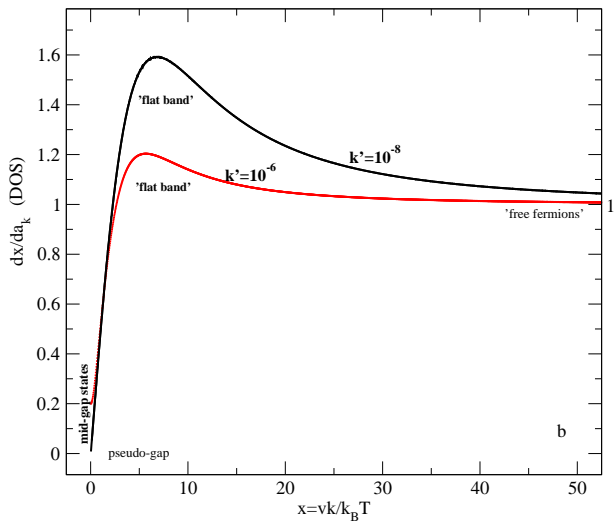}
\end{center}
\caption{Density of states (DOS) for the same values of $k'$ as in Fig. 1.}
\label{2}
\end{figure}

\noindent
As it follows from Eq. (\ref{aspec1}), when substituting into it Eq. (\ref{kprime}), the density of states in the center of the PG grows with decreasing coupling strength $|g|\approx U$ (and thus increasing parameter $k'$ in Eq. (\ref{kprime})), while the PG amplitude decreases simultaneously. When combined with Hall effect data in cuprates discussed in \cite{gorkov}, the in-gap states and PG behavior found in the present work may be tolerated qualitatively with the PG behavior inferred from Hall experiments if we assume that couplings $|g|,U$ {\it{decrease}} with the hole doping of the high-T$_c$ cuprates. Finally, we speculate, that if the instanton contribution to the partition path-integral Eq. (\ref{Z}) becomes negligable with respect to the contribution of the static SC or SDW order when the temperature $T$ lowers, the ground state is either superconducting or antiferromagnetic. On the other hand, another scenario to
be considered is when the system freezes while remaining in the instanton sector of its Hilbert space. The set of parameters that enables such behavior will be investigated in the forthcoming work.  
\\
\noindent
The author appreciates stimulating discussions with the colleagues at the Lorentz Institute for Theoretical Physics in Leiden and acknowledges financial support from the RFBR grant 07-02-12058 by Russian academy of sciences.
\\
*email address: sergeimoscow@online.ru

\end {document}